\documentclass[sigconf,screen]{acmart}
\AtBeginDocument{%
  \providecommand\BibTeX{{%
    \normalfont B\kern-0.5em{\scshape i\kern-0.25em b}\kern-0.8em\TeX}}}

\copyrightyear{2026}
\acmYear{2026}
\setcopyright{cc}
\setcctype{by}
\acmConference[FSE Companion '26]{34th ACM Joint European Software Engineering Conference and Symposium on the Foundations of Software Engineering}{July 05--09, 2026}{Montreal, QC, Canada}
\acmBooktitle{34th ACM Joint European Software Engineering Conference and Symposium on the Foundations of Software Engineering (FSE Companion '26), July 05--09, 2026, Montreal, QC, Canada}
\acmDOI{10.1145/3803437.3805556}
\acmISBN{979-8-4007-2636-1/2026/07}

\usepackage{verbatim}
\usepackage{lscape}
\usepackage{multirow}
\usepackage{dirtytalk}
\usepackage{makecell}
\usepackage{svg}
\usepackage[caption=false]{subfig}
\graphicspath{{Images/}}

\begin{document}

\title{Advancing Evidence-Based Social Sustainability in Software Engineering: A Research Roadmap}

\author{Bimpe Ayoola}
\affiliation{%
  \institution{Dalhousie University}
  \city{Halifax}
  \country{Canada}
}
\email{bimpe.ayoola@dal.ca}

\author{Anielle Andrade}
\affiliation{%
  \institution{Federal University of Pampa}
  \city{Alegrete}
  \country{Brazil}
}
\email{anielleandrade@unipampa.edu.br}

\author{Ronnie de Souza Santos}
\affiliation{%
  \institution{University of Calgary}
  \city{Calgary}
  \country{Canada}
}
\email{ronnie.desouzasantos@ucalgary.ca}

\author{Paul Ralph}
\affiliation{%
  \institution{Dalhousie University}
  \city{Halifax}
  \country{Canada}
}
\email{paulralph@dal.ca}

\renewcommand{\shortauthors}{Ayoola et al.}

\setlength{\fboxsep}{6pt}

\newcommand{\ResearchChallenge}[2]{%
  \par\smallskip
  \noindent\fbox{%
    \parbox{\dimexpr\columnwidth-2\fboxsep-2\fboxrule\relax}{%
      \textbf{#1}\ #2%
    }%
  }%
  \par\smallskip
}

\begin{abstract}
Social sustainability in software development means creating and maintaining systems that promote pro-social values (e.g., human well-being, equity), both now and in the future. However, social sustainability lacks clear conceptual and methodological foundations, and often takes a back seat to speed and profit. This paper therefore reports a narrative review of existing definitions of social sustainability in software development and identifies key aspects of social sustainability including social equity, well-being, and community cohesion. Challenges around measuring and integrating social sustainability into practice are conceptually analyzed. The paper then proposes a comprehensive definition of social sustainability and outlines a roadmap for measuring and integrating social sustainability into software engineering processes.
\end{abstract}



\maketitle

\section{Introduction}\label{sec:introduction}

Despite how software technology has transformed society across nearly every domain of human activity, software engineering (SE) research has historically focused on technical outcomes such as performance, reliability, and security, often at the expense of broader considerations about societal impact~\cite{al2014social,hussain2020human,mcguire2023sustainability}.

Sustainability has emerged as a critical concern in SE. Development (of software, cities, machines, etc.) is \textit{sustainable} to the extent that it meets the needs of the present without compromising our (or our descendants) ability to meet our (or their) future needs~\cite{greene2009measuring}. More broadly, sustainability includes multiple dimensions: environmental, economic, social, and technical~\cite{mcguire2023sustainability}. While environmental and economic sustainability have received increasing attention, the social dimension remains under-researched~\cite{al2014social, gustavsson2020blinded, mcguire2023sustainability}.

This lack of systematic attention to social sustainability is problematic because software systems can marginalize vulnerable populations, fragment communities, reinforce inequalities, and degrade human well-being even when they perform well on technical metrics. Making social sustainability central to SE requires a scientific approach with clear definitions, valid measurement strategies, and systematic evaluation of interventions intended to improve social sustainability outcomes.

Social sustainability is difficult to study because it is value-laden, context-dependent, and operates across multiple levels from individuals and teams to communities and societies~\cite{mcguire2023sustainability}. Furthermore, there is no consensus on its definition, and existing definitions often conflate two targets: the social properties of software systems in use and the social conditions under which software is developed~\cite{calero2022software,volpato2019has}. This lack of conceptual clarity complicates operationalization and measurement, contributing to empirical findings that are fragmented and hard to compare~\cite{martino2025, garciasustainability}.

This paper addresses key gaps in how social sustainability is conceptualized, measured, and studied in software engineering. We synthesize existing definitions and propose a refined definition (Section \ref{definition}), examine measurement challenges (Section \ref{measuring}), and present a research roadmap (Section \ref{roadmap}).

\section{\label{definition}Defining Social Sustainability}
To ground our work, we conducted a conceptual review of how social sustainability has been defined across SE, sustainability research, and related fields (e.g. urban planning). The sustainability literature emphasizes intergenerational equity and the fulfillment of human needs \cite{brundtland1987report}. Within SE, definitions highlight themes such as intergenerational equity, social justice, inclusion, participation, community well-being, and long-term societal impact~\cite{becker2015sustainability,lago2015framing,mcguire2023sustainability}.

Promoting \textbf{social equity and inclusion} is a recurring theme. Definitions frequently highlight the need for software to foster fairness, equity, and inclusion across all user groups---regardless of their socioeconomic status or geographical location---to minimize the digital divide~\cite{koning2001social,mckenzie2004social,moises2023social, moises2023socialreview}. \textbf{Human well-being} is another significant theme, extending beyond basic functionality to encompass users' physical, mental, and social health. Socially sustainable software should meet user needs while being secure and promoting a higher quality of life~\cite{ghahramanpouri2013urban,condori2018characterizing}. \textbf{Community well-being and social cohesion} also appear frequently, framing social sustainability as the capacity of communities to thrive and adapt over time \cite{mckenzie2004social,littig2005social,lago2015framing}. In SE contexts, this translates into software that enhances rather than fragments communities, supports collective action, and fosters belonging~\cite{mckenzie2004social, becker2015sustainability, lago2015framing}.
\textbf{Developer well-being and working conditions} emerge as an additional theme in some SE definitions, though less frequently than user-focused themes. Some highlight the social conditions of software development, including working conditions, human dignity, participation in decision making, and professional well-being within development environments \cite{littig2005social,calero2022software}.

Beyond the SE, insights from urban literature offer additional perspectives on social sustainability. These studies emphasize harmonious social evolution, cultural diversity, and improved quality of life~\cite{polese2000social,yiftachel1993urban}. In these contexts, social sustainability is tied to social equity and the sustainability of community~\cite{dempsey2011social}, a just and equitable society capable of adapting to evolving risks~\cite{eizenberg2017social}, creating environments that promote peaceful coexistence, social integration, and improved quality of life for everyone~\cite{polese2000social}.

Although these themes provide valuable guidance, existing definitions of social sustainability in SE vary in scope and emphasis. Some definitions focus primarily on the social impacts of software systems: what software does to the world and how it affects users, communities, and society~\cite{koning2001social, moises2023social, ghahramanpouri2013urban, condori2018characterizing}. Others incorporate concerns about development practices and working conditions, combining both product and process considerations within a single definition~\cite{calero2022software, mckenzie2004social, littig2005social}. This variation reflects the breadth of social sustainability, but it can also introduce ambiguity about what exactly is being evaluated in a given context.

Such ambiguity has practical and scientific implications. For researchers, unclear conceptual boundaries make it difficult to operationalize social sustainability constructs, develop appropriate measurement instruments, or compare findings across studies. For practitioners, varying definitions create conflicting guidance about what social sustainability requires in software projects. 

To address these challenges, we synthesize key themes from existing definitions, including social equity, human well-being and quality of life, community cohesion, human dignity, participation, and developer well-being, into two definitions that address different aspects of social sustainability in SE. Unlike prior definitions that vary inconsistently in which constructs they include~\cite{mckenzie2004social, lago2015framing, moises2023social, koning2001social} or conflate product and process concerns~\cite{calero2022software, littig2005social}, treating social sustainability as a single unified concept, our two definitions explicitly separate socially sustainable software (addressing the impacts of software systems on users and society), and socially sustainable software development (the conditions under which software is created and maintained). 

We propose the following definition of socially sustainable software, focusing specifically on the social impacts of software systems in use.
\newtheorem{ssdefinition}{Definition}
\begin{ssdefinition}
\textbf{Socially Sustainable Software} refers to a system that equitably serves diverse users, advances social justice and inclusion, protects privacy and well-being, strengthens rather than fragments communities, respects autonomy, and creates long-term positive value for all stakeholders without shifting negative impacts to marginalized groups, future generations, or other sustainability dimensions.
\end{ssdefinition}

We propose the following definition of socially sustainable software development, focusing on the social conditions and practices under which software is created:
\begin{ssdefinition}
\textbf{Socially Sustainable Software Development} is the process of maintaining equitable and humane working conditions for all practitioners involved, enabling diverse participation in technical decisions, systematically incorporating methods for assessing and addressing social impacts, engaging meaningfully with affected stakeholders, and sustaining this process over time without degrading the dignity, well-being, or professional growth of those who create and maintain software.
\end{ssdefinition}

\section{\label{measuring}Measuring Social Sustainability}
Measuring social sustainability presents distinct challenges for researchers and practitioners. Social sustainability involves latent, value-laden constructs---such as equity and well-being---that cannot be observed directly in software artifacts~\cite{ralph2024teaching}. By contrast, other sustainability dimensions in SE benefit from more established measurement approaches. Environmental sustainability is commonly assessed using green software metrics such as energy consumption~\cite{lago2014systematic}, and technical sustainability from established approaches to software quality metrics and technical debt~\cite{perera2024systematic}. However, social sustainability lacks comparable observable indicators or standardized measurement frameworks~\cite{al2014social}. Three challenges complicate measurement.

First, what constitutes socially sustainable software varies across geographical, cultural, legal, and socioeconomic contexts~\cite{oyedeji2017sustainability}. A system that promotes equity in one region may be ineffective or harmful in another. Measurement frameworks that insufficiently attend to context risk producing misleading or ethically problematic interpretations.

Second, social sustainability manifests simultaneously at individual, team, organizational, and societal levels \cite{mcguire2023sustainability}. A system may appear socially sustainable at one level while causing harm at another. Without explicitly modeling levels, assessments may show positive outcomes at one level while masking adverse effects elsewhere, risking drawing invalid conclusions. 

Third, social sustainability involves constructs that resist straightforward operationalization~\cite{al2014social,lami2014iso}. Human well-being, equity, community cohesion, diversity, and quality of life are abstract and difficult to quantify or operationalize as simple metrics.

These three challenges explain why existing evidence remains fragmented. Usable measurement approaches must instead make explicit the target of evaluation, the level of analysis, the stakeholder groups involved, and the value assumptions they encode. Without this clarity, measurement risks becoming fragmented, inconsistent, or disconnected from the social outcomes it aims to improve.

\section{\label{roadmap}A Roadmap for Evidence-Based Measurement and Integration}

Advancing social sustainability requires coordinated progress across conceptual clarity, measurement, and empirical evaluation. This roadmap outlines a methodical and evidence-based approach to driving social sustainability within software development.

\subsection{Create Social Sustainability Interventions}
Moving beyond discussion, we need concrete, empirically-evaluated interventions: deliberate actions or strategies implemented within the software development process to achieve sustainability goals \cite{becker2015sustainability}. 
These interventions must address both targets identified earlier: socially sustainable software and socially sustainable software development. \textit{Training based interventions} can help developers anticipate social impacts by building skills such as empathy~\cite{cerqueira2023thematic, gunatilake2024enablers}, ethical responsibility~\cite{rashid2015managing}, and cultural sensitivity~\cite{jaakkola2012culture}, supporting more inclusive design decisions and awareness of downstream social effects. \textit{Nudge-based interventions} leverage subtle cues to influence developer behavior toward social sustainability by making sustainable options and impacts more salient at decision points, for example by flagging potential bias risks, highlighting accessibility concerns, or surfacing the perspectives of diverse user personas during design and implementation. \textit{Process-level interventions} formalize social sustainability checks within development workflows by introducing new development processes or modifying existing ones to incorporate social impact checks into workflows. This may involve incorporating a ``Sustainability Impact Assessment'' (SIA)~\cite{berger2010sustainability} into backlog prioritization, where development teams would evaluate each user story or task for its potential sustainability impact before it is added to the sprint backlog. \textit{Regulatory and policy interventions} play a crucial role in shaping sustainable software development practices across the industry. These interventions typically rely on laws, regulations, guidelines, and policies that mandate or incentivize social sustainable practices in software development. These policies can be formalized practices, standards, or rules that align teams and developers with broader sustainability goals~\cite{UN_SDGs}.


\ResearchChallenge{Research Challenge 1.}{Develop \textit{many} practical, sustainability promoting interventions for software professionals, teams, and organizations.}

\subsection{Create Measurement Infrastructure}
Progress on social sustainability requires credible, usable measurement approaches that align with what is being evaluated. Existing work shows that social sustainability lacks standardized metrics within SE~\cite{garciasustainability}, and importing measures from other disciplines is not straightforward~\cite{Brink2020}. While psychology and sociology or related fields offer validated instruments for constructs such as psychosocial well-being~\cite{hills2002oxford}, organizational inclusion~\cite{sauter1997psychometric}, and community engagement~\cite{waterton2014heritage}, these require substantial adaptation and validation for software contexts and may not be suitable for short term or lab based studies common in SE research. A comprehensive measurement model for social sustainability does not yet exist, and developing one requires validated measures for its constituent components. One promising direction is the development of targeted metrics for common social sustainability concerns, such as accessibility (\textit{user accessibility score}~\cite{song2017waem}), inclusivity (\textit{social inclusivity index}~\cite{gunaruwan2015social}), well-being (user well-being impact scale~\cite{stewart2008warwick}), or community impact(\textit{community engagement score}~\cite{goodman2017evaluating}), grounded in established standards where possible.

Surveys, interviews, and case studies can also supplement these quantitative metrics by providing qualitative insights into the software's impact on users, developers, stakeholders and communities. Measures should explicate whether they target individual experience, team working conditions, organizational practices, or societal outcomes. Without this clarity, findings will remain difficult to compare or synthesize across studies.

\ResearchChallenge{Research Challenge 2.}{Establish a standard measurement model for social sustainability construct.}

\subsection{Design Appropriate Intervention Studies} 

Most SE researchers create \textit{interventions} that ostensibly improve project outcomes when used by professional developers, but shy away from \textit{intervention studies}. Case studies, grounded theory, questionnaires, repository mining, nonexperimental simulation, and exploratory data science can all help establish feasibility and plausibility, especially given the cost of intervention studies. Longitudinal analysis or meta-analysis may also provide evidence when interventions are widely adopted. However, such methods alone cannot determine whether an intervention actually improves social sustainability outcomes in real development contexts. We need either action research (if we prefer a qualitative approach) or randomized, controlled experiments (if we prefer a quantitative approach).\footnote{We are glazing over quasi-experiments because they are only justified when random assignment is impractical, which should be rare in the circumstances we are discussing.} 

In other words, to empirically evaluate a social sustainability intervention~\cite{ralph2024teaching}, the appropriate dependent variable must target social sustainability outcomes, or a closely related concept; the appropriate hypothesis is that our invention improves the dependent variable; and the best method is either action design research~\cite{sein2011action} or a randomized, controlled experiment.

\subsubsection{Designing Smarter Experiments}

Experiments should use realistic tasks, authentic development environments, and decision contexts where social sustainability trade-offs naturally arise. Ideally, such studies would use stratified random sampling (see~\cite{baltes2022sampling}) to select companies and then teams, randomly assign these teams to either a control group or a treatment group, have them perform specific tasks, and measure social sustainability outcomes. In practice, this level of realism and sampling rigor is rarely feasible. Merely recruiting a reasonable-sized convenience sample of software professionals to participate in a lab-based study is nearly impossible because few, if any, granting agencies are willing to provide sufficient financial incentives.
We have three suggestions for overcoming the professional recruiting problem:

\begin{enumerate}
    \item Acknowledge that, in many cases, a convenience sample of SE (or computer science, etc.) students is a reasonable simplification, and certainly better than researchers testing their interventions themselves.
    \item Use a randomized block design~\cite{ariel2010randomized} combining undergraduate students, graduate students, and a smaller number of professionals. Instead of excluding participants with little experience, model experience as a covariate and empirically assess whether experience moderates the main effect.
    \item To recruit a modest number of professional participants, organize a social event in which participating in the study is one of several activities. Consider including a short keynote talk, panel discussion, tutorial, or ask-me-anything session. Consider working with an organization that frequently organizes such events (e.g. an incubator or meetup group), rather than going it alone.
    \item Avoid online recruiting or crowdsourcing services due to the risk of participant fraud (e.g. people pretending to have skills or experience they do not have, and using AI or other tools to complete experimental tasks for pay). Conduct experiments in-person, where participants can be observed.
\end{enumerate}

\subsubsection{Utilize Action Research}

Action research refers to a family of research methods that involve iterative cycles of 
making sense of a site, planning an intervention, intervening, assessing outcomes, and reworking the intervention; adopting a critical philosophical 
position that research aims to improve our world; and primarily 
qualitative data collection and analysis. Several variations on action research have been proposed including Action Design Research (ADR)~\cite{sein2011action}, which combines action research with building innovative artifacts; and participatory action research (PAR)~\cite{whyte1991participatory}, which emphasizes including people affected by research as partners (not subjects).  ADR and PAR are good choices for assessing sustainability interventions because they: (1) take place in real organizations, with real developers, doing real projects; (2) emphasize theory development; (3) do not require a sophisticated, a priori, quantitative measurement model; (4) are naturally longitudinal, which matches the long-term effects of sustainability; and (5) have well-understood applicability to SE research~\cite{staron2020action}. 

\ResearchChallenge{Research Challenge 3.}{Evaluate social sustainability interventions using rigorous randomized, controlled experiments and Action Design Research (ADR)}

\subsection{Facilitate Interdisciplinary Collaboration and Longitudinal Social Sustainability Research}
Social sustainability sits at the intersection of SE, social science, ethics, and user-centered design. Interdisciplinary collaboration is essential for defining constructs, validating instruments, and engaging affected stakeholders in meaningful ways. Partnerships across industry, academia, and government can accelerate the development of social sustainability metrics and tools. While researchers provide the theoretical underpinnings and innovative methodologies for measuring social sustainability, engaging industry partners will help researchers ensure that the metrics are applicable in real-world scenarios.  In addition, many social sustainability outcomes emerge over long time horizons. Capturing cumulative effects on wellbeing, labor conditions, or community cohesion requires multi-year longitudinal studies, which are difficult to fund under typical grant cycles and 
challenging for junior researchers building publication records. Addressing this gap requires greater support for interdisciplinary and longitudinal research designs for investigating long-term social outcomes.

\ResearchChallenge{Research Challenge 4.}{Support interdisciplinary collaboration and multi-year longitudinal studies to evaluate long term social sustainability outcomes.}
\section{\label{conclusion}Conclusion}
Social sustainability represents a critical yet underexplored dimension of SE. While environmental and economic sustainability have gained increasing attention, the social dimension remains poorly defined, inadequately measured, and rarely evaluated through rigorous empirical research.

This paper makes three main contributions: (1) a novel, comprehensive definition of social sustainability in the SE context, emphasizing the difference between socially sustainable software and socially sustainable software development (Section \ref{definition}); (2) a discussion of measurement challenges (Section \ref{measuring}), highlighting that social sustainability involves latent and value-laden constructs that vary across contexts and operate at multiple levels of analysis; (3) a research roadmap (Section \ref{roadmap}) centered on four imperatives: developing empirically evaluated interventions, establishing credible and standardized measurement infrastructure, designing more rigorous experimental studies to evaluate social sustainability interventions, and fostering interdisciplinary collaboration and longitudinal research. Realizing this agenda will require sustained effort and coordination across research communities, funding bodies, and industry partners. We also acknowledge that this paper is primarily conceptual and reflects the perspectives of its authors and our understanding of the field's needs. 

As software increasingly mediates access to resources, opportunities, and social participation, ensuring that it promotes equity, human well-being, and long-term societal value should be a central research priority. Preventing social harm warrants at least as much attention as optimizing technical performance in SE research. This paper aims to provide a foundation for advancing social sustainability as an evidence-based practice in software engineering and to encourage the community to invest sustained research effort in this neglected but essential area.

\begin{acks}
    This study was supported by the National Sciences and Engineering Research Council of Canada (RGPIN-2020-05001).
\end{acks}

\bibliographystyle{ACM-Reference-Format}
\bibliography{references}

\end{document}